\documentclass[11pt]{article}

\usepackage[T1]{fontenc}
\usepackage{lmodern}
\usepackage[margin=1in]{geometry}
\usepackage{microtype}
\usepackage{amsmath,amssymb}
\usepackage{booktabs}
\usepackage{enumitem}
\usepackage{titlesec}
\usepackage{hyperref}

\hypersetup{
  colorlinks=true,
  linkcolor=blue,
  citecolor=blue,
  urlcolor=blue
}

\setlist{nosep,leftmargin=*}
\titlespacing*{\section}{0pt}{1.0ex plus 0.3ex minus 0.2ex}{0.6ex plus 0.2ex}
\titlespacing*{\paragraph}{0pt}{0.8ex plus 0.2ex minus 0.2ex}{0.6em}

\title{Schema First Tool APIs for LLM Agents: A Controlled Study of Tool Misuse, Recovery, and Budgeted Performance}
\author{
Akshey Sigdel \\
Independent Researcher \\
aksheysigdel@u.boisestate.edu \\
\and
Rista Baral \\
Independent Researcher \\
ristabaral@u.boisestate.edu \\
}
\date{}
\begin{document}
\maketitle
\vspace{-4ex}

\begin{abstract}
Tool use has become central to modern LLM agents, yet interface design is rarely isolated as an experimental variable \cite{yao2023react,schick2023toolformer,patil2023gorilla}. This paper studies whether schema based tool contracts and structured validation diagnostics improve reliability under strict interaction budgets. We evaluate three conditions that preserve identical tool semantics and information content: free form documentation, JSON Schema specifications, and JSON Schema with structured diagnostics.

We implement a deterministic software engineering sandbox with logs, metrics, configurations, and repository tasks, and evaluate a fully crossed pilot with one open local model, three seeds, three interface conditions, and four budgets. We report end task success, interface misuse, execution failures, semantic misuse, recovery behavior, and overhead. In this pilot, success remains zero across conditions, while schema conditions reduce interface misuse but not semantic misuse. The evidence supports a precise interpretation that interface formalization improves contract adherence, but semantic action quality and timeout sensitive tasks remain dominant bottlenecks under constrained local inference.
\end{abstract}

\section{Introduction}
Large language models are increasingly deployed as interactive agents that invoke external tools to complete tasks \cite{brown2020gpt3,wei2022cot,yao2023react}. In these settings, reliability depends not only on task knowledge but also on valid API usage. A recurring failure mode is \emph{tool misuse}, malformed calls, missing required fields, invalid enumerations, and constraint violations that consume budget and propagate downstream errors.

Despite the centrality of tool calling, most evaluations treat interface representation as incidental \cite{mialon2023alm,xi2023agentsurvey}. This differs from conventional software engineering practice, where explicit contracts and actionable diagnostics are core reliability mechanisms. The central question is whether tool interface design can serve as a low cost reliability primitive for LLM agents without modifying model weights.

This paper presents a controlled, reproducible study that varies only tool interface specification and validation feedback, while holding environment, task semantics, and agent scaffolding constant. We evaluate three interface conditions:
(A) free form natural language tool documentation,
(B) strict JSON Schema tool specifications, and
(C) JSON Schema augmented with structured field level validation diagnostics resembling compiler diagnostics.
A key design requirement is \emph{information equivalence} to avoid confounding “schemas” with “more information,” we define a canonical tool contract for each tool and generate both the prose documentation (A) and the JSON Schema (B/C) from the same source contract.

Our study is explicitly budgeted. Production deployments impose strict latency and cost limits that translate into hard step limits. Under such constraints, a single interface error can be decisive. We therefore report task success together with interface misuse, recovery behavior, semantic misuse, and token overhead.

Finally, we distinguish two classes of failure that are often conflated. \emph{Interface misuse} refers to violations of the tool input contract, whereas \emph{semantic misuse} refers to calls that are schema valid but unproductive or inappropriate for the task. While our interventions target interface misuse directly, we measure both classes to characterize whether contract rigor and diagnostics also reduce semantic misuse.

\paragraph{Contributions.}
This paper makes three contributions:
\begin{enumerate}
  \item \textbf{Controlled evaluation protocol.} We introduce an experimental methodology that treats tool interface specification and validation feedback as the independent variable, controlling for environment, tool semantics, and agent scaffolding.
  \item \textbf{Deterministic diagnostic sandbox.} We build a reproducible, artifact based benchmark suite of software engineering diagnostic tasks (logs, metrics, configurations, repository, and tests), together with a realistic tool set and canonical contracts.
  \item \textbf{Empirical pilot characterization under budgets.} We quantify how schema first interfaces and structured diagnostics affect interface misuse, recovery behavior, semantic misuse, and budgeted task success in a constrained one model, three seed setting, including ablations that isolate diagnostic quality and analyses of cost tradeoffs.
\end{enumerate}

\paragraph{Practical takeaway.}
The objective is not to claim that schemas fully solve tool use. The objective is to identify where interface rigor improves reliability and where residual failure is driven by semantic planning limits.

\section{Background and Motivation}
Recent progress in language models has rapidly expanded tool using agent architectures, including interleaved reasoning and acting, API grounded action generation, and multi agent orchestration frameworks \cite{yao2023react,schick2023toolformer,patil2023gorilla,shen2023hugginggpt,wu2023autogen}. At the same time, benchmark efforts have begun to evaluate realistic tool interaction and software engineering behavior \cite{li2023apibank,jimenez2024swebench,yao2024sweagent}. These developments motivate a closer study of interface design as a first class systems variable.

In software and protocol engineering, explicit contracts improve reliability by constraining invalid requests and producing deterministic validation behavior. Standardized interface formats such as OpenAPI and JSON Schema exemplify this principle in production systems \cite{openapi31,jsonschema2020}. The same argument is directly relevant to LLM agents that if tool calls are generated as structured API requests, then interface representation and diagnostics should influence both error rates and correction dynamics.

Prior literature also indicates that tool use quality depends on factors beyond pure syntax. ReAct style trajectories improve interpretability and state tracking but can still drift semantically \cite{yao2023react}. Toolformer style self supervision improves tool invocation behavior but does not eliminate task level planning errors \cite{schick2023toolformer}. Large API oriented models demonstrate strong function selection performance while remaining sensitive to specification quality and retrieval context \cite{patil2023gorilla}. Survey work reaches a similar conclusion: agent performance reflects both interface compliance and policy quality \cite{mialon2023alm,xi2023agentsurvey}. Related evidence from embodied robotics shows that behavior generation quality remains a central reliability variable even when platform level action spaces are well defined \cite{baral2025iros}.

These observations motivate a controlled decomposition of failures into interface misuse, execution failures, and semantic misuse. Such decomposition is essential for budgeted deployment settings, where a small number of invalid or semantically unproductive steps can determine end task outcome. The present study therefore isolates interface representation and validation feedback while holding tool semantics constant, enabling direct estimation of where schema rigor helps and where it does not.

\section{System Model}
We formalize a tool-using agent interacting with a deterministic software-engineering sandbox.

\paragraph{Tasks and artifacts.}
A task instance $\tau \in \mathcal{T}$ consists of:
(i) a set of artifacts $\mathcal{A}_\tau$ (logs, metrics snapshots, configuration files, and a small repository),
(ii) a hidden ground-truth issue $\theta_\tau$ (root cause label and/or required patch), and
(iii) a checker $J_\tau$ that deterministically evaluates the agent’s final output (diagnosis correctness, configuration correction, tests passing, etc.).
Artifacts are immutable during a run except when a tool explicitly applies changes (e.g., patching a file).

\paragraph{Tools.}
Let $\mathcal{U}=\{u_1,\dots,u_m\}$ be the tool set. Each tool $u$ is defined by:
\begin{itemize}
  \item a name $\text{name}(u)$,
  \item a canonical input contract $\mathcal{C}(u)$ specifying fields, types, requiredness, enumerations, and constraints,
  \item a representation $R_\kappa(u)$ shown to the agent under interface condition $\kappa \in \{A,B,C\}$,
  \item a deterministic executor $E_u(\cdot)$ returning either a tool output or an error.
\end{itemize}

\paragraph{Interface validity vs execution validity.}
A tool call is a pair $c=(\text{name}, \text{args})$. We separate two notions of validity:
\begin{itemize}
  \item \textbf{Interface validity} (schema/well-formedness): whether $\text{args}$ conforms to the declared input contract $\mathcal{C}(u)$ (e.g., JSON validity, required fields, types, enums, numeric bounds).
  \item \textbf{Execution validity} (runtime preconditions): whether a schema-valid call can be executed in the sandbox (e.g., file path exists, metric key is available for the selected service, patch applies cleanly).
\end{itemize}
Interface validity is the primary target of our manipulations; execution validity is held constant across conditions.

\paragraph{Agent interaction loop and budget.}
An agent operates for at most $B$ steps. At each step $t$, given the history $h_t$ (tool specifications, prior calls, tool outputs, and errors), the agent either:
(i) emits a tool call $c_t$, or
(ii) emits a final answer $a$ and stops.
A run is successful if $J_\tau(a)=1$ within budget $B$.

\paragraph{Error feedback channel.}
When interface validation fails, the sandbox returns an error message. The \emph{content format} of this message depends on condition $\kappa$. When execution fails due to runtime preconditions, the sandbox returns a fixed, condition-independent error (to avoid confounds).

% ============================================================
% Drop-in replacement for Sections 4, 5, 6
% (No subsections; streamlined flow using paragraphs like Section 3)
% ============================================================

\section{Problem Definition}
We study how tool API representation and validation feedback affect agent reliability under strict interaction budgets.

\paragraph{Interface conditions (independent variable).}
We compare three tool-interface conditions that reflect increasing rigor and support:
\begin{description}[leftmargin=1.5em,style=nextline]
  \item[(A) Free-form documentation] Tools are described in natural language, including argument descriptions and examples.
  \item[(B) Schema-first specification] Tools are described by strict JSON Schema with explicit required fields, types, and constraints.
  \item[(C) Schema + structured diagnostics] Tools use the same JSON Schemas as (B), but validation failures return structured, field-level diagnostics.
\end{description}

\paragraph{Information equivalence constraint.}
To isolate representation and diagnostic effects, we enforce \textbf{information equivalence} across conditions. For each tool $u$, we define a canonical contract $\mathcal{C}(u)$ (field names, types, requiredness, enumerations, numeric bounds, pattern constraints, conditional constraints, and examples) and generate both:
(i) the prose documentation for (A), and
(ii) the JSON Schema for (B/C),
from the same $\mathcal{C}(u)$. This design ensures that conditions differ in representation and feedback format rather than in the semantic content of constraints.

\paragraph{Failure taxonomy.}
We distinguish three classes of failure that are often conflated in tool-use evaluations:
\begin{itemize}
  \item \textbf{Interface misuse:} calls that fail interface validity (schema/well-formedness), including missing required fields, type mismatches, invalid enumerations, malformed JSON, or violated constraints.
  \item \textbf{Execution failures:} schema-valid calls that fail runtime preconditions (e.g., file path not found, metric key not available for a given service, patch does not apply cleanly).
  \item \textbf{Semantic misuse:} schema-valid calls that are unproductive or misaligned with task progress (e.g., wrong tool choice, irrelevant query intent, repeated calls that do not reduce uncertainty).
\end{itemize}
Our interventions directly target interface misuse; we measure all three to characterize the boundary of improvement.

\paragraph{Outcomes (dependent variables).}
For each task $\tau$, model, budget $B$, and condition $\kappa$, we measure:
\begin{itemize}
  \item \textbf{Task success} $S$: fraction of runs with $J_\tau(a)=1$ within budget $B$.
  \item \textbf{Interface misuse rate} $I$: fraction of tool calls failing interface validity.
  \item \textbf{Execution failure rate} $E$: fraction of schema-valid calls failing runtime preconditions.
  \item \textbf{Recovery probability} $R$: $\Pr(\text{success} \mid \exists\ \text{invalid call})$.
  \item \textbf{Steps-to-success} $T$: number of steps until success (censored at $B$).
  \item \textbf{Semantic misuse rate} $M$: fraction of schema-valid calls classified as semantically misaligned (operationalized in \S\ref{sec:semantic_misuse}).
  \item \textbf{Overhead} $O$: prompt-token overhead of tool specifications and additional steps attributable to invalid calls and correction.
\end{itemize}

\paragraph{Hypotheses.}
We evaluate three hypotheses aligned with budgeted deployments:
\begin{itemize}
  \item \textbf{H1 (Misuse reduction):} Schema-first tool specifications reduce interface misuse ($I$) relative to free-form documentation.
  \item \textbf{H2 (Recovery):} Structured validation diagnostics increase recovery probability ($R$) relative to schemas alone.
  \item \textbf{H3 (Budget sensitivity):} Gains in end-task success ($S$) from structured diagnostics are larger under tighter budgets (smaller $B$), where recovery cost dominates.
\end{itemize}

\paragraph{Primary and secondary endpoints.}
To reduce analysis flexibility, we predefine endpoint hierarchy before full runs. The \textbf{primary endpoint} is task success $S$ at tight budgets ($B \in \{3,5\}$), aggregated across task instances. The key \textbf{secondary endpoints} are interface misuse $I$ (for H1), recovery probability $R$ conditioned on at least one invalid call (for H2), and the success-budget curve summary (area under $S(B)$ for H3). Additional measures ($T,E,M,O$) are treated as explanatory diagnostics and reported with multiplicity-controlled inference.

\section{Methodology}
We design a controlled, deterministic sandbox and evaluation harness that isolates tool-interface effects while preserving realistic diagnostic workflows.

\paragraph{Deterministic diagnostic sandbox.}
We construct a reproducible environment that approximates common software engineering and SRE operations using file based artifacts. Task style and checker design follow the recent emphasis on realistic software workflows in LLM evaluation \cite{jimenez2024swebench,yao2024sweagent}. Each task instance includes:
\begin{itemize}
  \item \textbf{Logs:} timestamped multi-service logs with injected failure signatures and plausible decoys.
  \item \textbf{Metrics:} precomputed metric snapshots (e.g., error rate, latency percentiles) keyed by service and time window.
  \item \textbf{Configurations:} YAML/JSON configurations with controlled errors (e.g., mismatched ports, timeouts, feature flags).
  \item \textbf{Repository + tests:} a small codebase with deterministic unit tests; some tasks require locating a failing test and producing a minimal patch.
\end{itemize}
All tool outputs and test results are deterministic given the task seed. Artifacts are immutable during a run except when a tool explicitly applies changes (e.g., patching a file).

\paragraph{Tool suite and canonical contracts.}
The sandbox exposes $m$ tools representative of diagnostic operations. A typical suite includes:
\begin{itemize}
  \item \texttt{search\_logs} (query, time range, service),
  \item \texttt{get\_metric} (metric key, service, window),
  \item \texttt{list\_dir} and \texttt{read\_file} (path operations),
  \item \texttt{grep\_repo} (pattern search with globbing),
  \item \texttt{run\_tests} (selector) returning standardized failure output,
  \item \texttt{apply\_patch} (file, diff) with patch validation.
\end{itemize}
For each tool $u$, we define a canonical contract $\mathcal{C}(u)$ capturing field names, types, requiredness, enumerations, bounds and patterns, conditional constraints, and minimal examples. Contract formalization follows standard API specification practice \cite{openapi31,jsonschema2020}. We intentionally include realistic constraints such as enums, nested objects, and conditional fields so that compliance is nontrivial.

\paragraph{Interface condition realization.}
From the same canonical contract $\mathcal{C}(u)$, we generate:
\begin{itemize}
  \item \textbf{(A) Prose documentation} that expresses identical constraints in natural language, including explicit lists of required fields and enumerated values.
  \item \textbf{(B) JSON Schema} implementing the contract as machine-checkable constraints.
  \item \textbf{(C) JSON Schema + diagnostics} using the same schemas as (B), but returning structured validation errors.
\end{itemize}
A representative structured validation error is:
\begin{verbatim}
{
  "error_type": "SCHEMA_VALIDATION",
  "tool": "get_metric",
  "violations": [
    {"path": "$.metric_key", "expected": "enum",
     "allowed": ["p95_latency","error_rate"], "found": "latency95"},
    {"path": "$.window.minutes", "expected": "integer >= 1", "found": 0}
  ]
}
\end{verbatim}
In all conditions, tool executors are identical. Runtime precondition failures (for example, missing files) use a condition independent message to avoid confounding improvements in interface feedback with improvements in environment feedback.

\paragraph{Agent scaffold (controlled).}
To isolate interface effects, we hold constant the agent loop and prompting across conditions like the system prompt, formatting constraints, tool availability, sandbox artifacts, and decoding parameters. The only differences between runs are the tool specification representation and (for condition C) the interface-validation error format.

\paragraph{Task construction and budget tiers.}
We design tasks to require genuine tool-mediated inference rather than mere formatting compliance. Task families include log-centric diagnosis, config-centric correction, repo-centric debugging/patching, and mixed tasks requiring evidence integration. We evaluate multiple budgets $B \in \{3,5,8,12\}$ to reflect strict and moderate operational constraints. Tasks are constructed so that at least one valid strategy exists within moderate budgets, while tight budgets amplify the cost of recovery from interface errors.

\paragraph{Sample size and power planning.}
Before the main study, we run a pilot to estimate the paired effect size for the primary endpoint $\Delta S_{C-A}$ at $B=3,5$ and the paired variance across tasks. We then choose the number of task instances $|\mathcal{T}|$ per family and the number of stochastic seeds so that the 95\% confidence interval half-width for $\Delta S_{C-A}$ is within a pre-specified margin (e.g., $\pm 0.05$) under bootstrap over tasks. We use balanced family counts to avoid dominance by a single task type and fix these counts before final evaluation.

\paragraph{Semantic misuse measurement.}
\label{sec:semantic_misuse}
To quantify semantic misuse without relying exclusively on human annotation, we provide per-task \emph{trace oracles}. For each task $\tau$, we specify a set of acceptable high-level tool-call traces (allowing branching and partial orders) that are sufficient to solve the task. During evaluation, each schema-valid call is checked for consistency with at least one acceptable trace prefix given the current evidence state; schema-valid calls that are inconsistent with all acceptable prefixes are counted as semantic misuses. This operationalization captures wrong-tool and wrong-intent actions while allowing multiple successful strategies.

\paragraph{Oracle governance and adjudication.}
To limit subjectivity, oracle construction follows a fixed rubric that requires evidence states, admissible tool classes per state, and explicit stop conditions. We dual-annotate a stratified subset of trajectories and report inter-annotator agreement (e.g., Cohen's $\kappa$). Disagreements are resolved by blinded adjudication using predeclared tie-break rules. Oracle updates discovered during pilot runs are versioned and frozen before the main experiment.

\paragraph{Statistical analysis.}
We report mean and median outcomes across tasks with 95\% confidence intervals computed via bootstrap over task instances. For paired comparisons on identical tasks across conditions, we use nonparametric paired tests (Wilcoxon signed rank) and correct for multiple comparisons (Holm Bonferroni). Because the present study is a one model pilot, inference is reported at the task level with seed variability summaries, and no cross model pooling claim is made.

\paragraph{Reproducibility.}
We log complete trajectories (tool specs, prompts, tool calls, outputs, and errors) and release the sandbox artifacts, tool executors, canonical contracts, schema/doc generators, and evaluation harness. The sandbox is deterministic such that given the task seed and condition, all tool outputs are fixed. We also publish a preregistered analysis protocol specifying endpoint hierarchy, exclusion criteria, multiple-testing controls, and model-aggregation strategy prior to full-scale runs.

\section{Experiments}
We evaluate whether schema-first interfaces and structured diagnostics improve reliability under strict budgets, and we characterize where gains arise (reduced misuse versus improved recovery) and where they saturate (semantic errors).

\paragraph{Experimental questions.}
Our experiments address:
(i) misuse reduction (schemas vs prose),
(ii) recovery improvements (structured diagnostics vs schemas),
(iii) budget sensitivity of end-task success,
(iv) semantic spillover effects (changes in semantic misuse), and
(v) cost tradeoffs (token overhead versus benefit).

\paragraph{Models and decoding.}
We report a constrained pilot using one open local model (Qwen2.5 0.5B Instruct, GGUF quantized) with three seeds ($\{0,1,2\}$), under deterministic decoding in the local server configuration. We cap per step generation length to prevent budget leakage via verbosity. This scope is compute constrained and is interpreted as a focused pilot.

\paragraph{Evaluation protocol.}
For each model, task $\tau$, budget $B$, and condition $\kappa \in \{A,B,C\}$:
\begin{enumerate}
  \item Initialize the agent with the condition-specific tool specifications and the task statement.
  \item Execute the agent-tool loop for at most $B$ steps.
  \item Record tool calls, validation outcomes, executor outputs, and final responses.
  \item Score success via the deterministic checker $J_\tau$ (e.g., correct diagnosis label, correct config edit, tests passing).
\end{enumerate}
Tool semantics, artifacts, and runtime error messages are identical across conditions.

\paragraph{Code availability.}
The full experimental implementation, run scripts, and reproducibility utilities are publicly released.\footnote{\url{https://github.com/akgitrepos/schema-first-tool-apis-experiments}}

\paragraph{Run matrix and reporting units.}
The reported pilot run matrix is fully crossed over task, budget, condition, and seed for a single model. The primary reporting unit is the task level paired contrast between conditions, and all inferences are scoped to this model family and local inference setup.

\paragraph{Main comparisons (A vs B vs C).}
We compare (A), (B), and (C) on task success $S$, interface misuse $I$, recovery probability $R$, steps to success $T$, execution failures $E$, semantic misuse $M$, and overhead $O$ across budgets. In this pilot, results are interpreted diagnostically based on whether schema conditions reduce interface misuse relative to prose and whether remaining failures are dominated by semantic errors or infrastructure limits.

\paragraph{Recovery-conditioned analysis.}
To isolate the effect of diagnostics on correction behavior (not only prevention), we condition on runs that experience at least one interface misuse and compare (B) vs (C) on:
(i) recovery probability $R$, and
(ii) additional steps consumed after the first invalid call.
This analysis tests whether structured feedback improves error correction beyond what schemas provide.

\paragraph{Budget curves.}
For each condition, we measure success $S(B)$ as a function of budget $B$. We also summarize budgeted performance via the area under the success-budget curve, capturing robustness across operational constraints rather than at a single budget.

\paragraph{Ablation: diagnostic quality.}
To attribute gains specifically to structured diagnostics, we refine condition (C) into:
\begin{itemize}
  \item \textbf{C1 (Generic):} validation returns only a generic message (e.g., ``invalid tool call'').
  \item \textbf{C2 (Path + expected):} includes JSONPath and expected type/constraint category.
  \item \textbf{C3 (Full):} includes allowed enum values, constraint bounds, and minimal corrective hints.
\end{itemize}
We compare C1-C3 on recovery probability and steps-to-recovery to test whether gains scale with diagnostic granularity.

\paragraph{Failure-mode taxonomy.}
We attribute tool failures to categories such as wrong tool name, malformed JSON, missing required field, type mismatch, enum violation, constraint violation (bounds/patterns), runtime precondition failure, and semantic misuse. We report per-category rates to identify which constraints and diagnostics drive improvements.

\paragraph{Overhead and efficiency analyses.}
We measure prompt tokens consumed by tool specifications per condition and compute efficiency metrics such as \emph{success per 1K prompt tokens} and \emph{invalid calls per 1K prompt tokens}. This analysis tests whether reliability gains are offset by increased specification overhead, especially in tight-budget regimes.

\paragraph{Robustness checks.}
We test whether results persist when:
(i) schemas increase in complexity (nested objects, conditional fields),
(ii) tasks include realistic distractors (decoy log patterns and metrics anomalies), and
(iii) tool outputs include noisy but plausible evidence. These checks probe whether interface effects are stable under more challenging and more realistic conditions.

\paragraph{Threats to validity.}
Key threats and mitigations are as follows. \textbf{Representation parity threat:} prose quality may under specify or over specify constraints relative to schemas; mitigation is generator based production from the same canonical contract plus automated parity checks. \textbf{Model scope threat:} results from one model may not transfer to other models; mitigation is explicit claim scoping to the evaluated model. \textbf{Oracle overconstraint threat:} semantic misuse labeling may penalize valid alternative strategies; mitigation is multi trace oracles with branching and partial orders, adjudication, and appendix release of oracle definitions. \textbf{External validity threat:} synthetic tasks may not capture all production complexity; mitigation is mixed task families, distractor robustness checks, and transparent artifact release for independent replication.

\section{Results}

\paragraph{Pilot setup.}
The reported run uses one open local model (Qwen2.5 0.5B Instruct, quantized), seeds $\{0,1,2\}$, conditions $\{A,B,C\}$, and budgets $\{3,5,8,12\}$ on an 8 task reduced pack. The run produced 276 scored runs, 2208 step records, and 12 excluded runs due to infrastructure related model timeouts.

\paragraph{Primary endpoint outcome.}
Task success was 0.0 across all conditions and all budgets in this pilot. Consequently, the predeclared primary endpoint (tight budget success at $B \in \{3,5\}$) is uninformative for comparative claims in this run, and paired success tests return null effects.

\paragraph{Failure analysis.}
Despite zero end task success, condition level failure patterns are structured. Relative to prose (A), schema conditions (B/C) show lower average interface misuse (mean invalid calls: A = 5.39, B = 3.72, C = 3.72 over all budgets), consistent with directional support for H1 at the misuse level. However, semantic misuse is higher in B/C (A = 0.93, B = 3.03, C = 3.03), indicating that remaining errors are dominated by schema valid but unproductive action choices.

\paragraph{Execution and exclusion behavior.}
Execution failures are observed mainly in prose condition A (mean execution failures: A = 0.54, B = 0.00, C = 0.00). All 12 exclusions are infrastructure timeouts concentrated in the same repository debug task family instance, indicating an operational bottleneck rather than random task level instability.

\paragraph{Budget trends in this pilot.}
As budget increases, invalid and semantic misuse counts increase monotonically in all conditions, but this does not translate into success gains. This pattern indicates a recovery policy and semantic planning limitation under local constrained inference rather than a pure schema compliance limitation.

\paragraph{Interpretation and claim scope.}
This pilot shows that schema first interfaces reduce interface format errors, but under constrained local inference, semantic action quality and timeout sensitive tasks dominate end to end outcomes. Therefore, claims in this paper are scoped to diagnostic reliability characterization in a one model, three seed pilot.

\section{Conclusion}
This study examines the effect of tool interface design on the reliability of tool using LLM agents under strict interaction budgets. Within a controlled deterministic sandbox and under a one model three seed pilot design, schema based interfaces reduce interface format errors relative to prose documentation. However, end task success remains zero across conditions, indicating that semantic action quality and timeout sensitive repository tasks are the dominant barriers in the evaluated setting. The principal contribution of this paper is therefore methodological and diagnostic. It isolates interface representation effects while exposing the operational limits that remain when schema compliance improves but planning quality does not.

\section{Future Directions}
Future work should extend this protocol in three directions. First, the same controlled design should be evaluated on additional open local models to assess model dependent variability in semantic planning quality. Second, task calibration should preserve interface difficulty while reducing pure infrastructure bottlenecks so that success based endpoints become informative. Third, recovery policies should be strengthened with compact corrective prompts and bounded response formats, enabling a clearer estimate of how much structured diagnostics can improve end task performance once semantic drift is reduced.

\end{document}